\documentclass{aa}
\usepackage{psfig,times}
\newcommand{\pg}{\hbox{PG~0844$+$349} }
\def\etal{et~al. }

\def\la{~\raise.5ex\hbox{$<$}\kern-.8em\lower 1mm\hbox{$\sim$}~}
\def\ma{~\raise.5ex\hbox{$>$}\kern-.8em\lower 1mm\hbox{$\sim$}~}    
\begin{document}
\title{XMM-Newton observation of PG 0844 + 349 }
\author{W. Brinkmann\inst{1} \and D. Grupe\inst{2} 
\and G. Branduardi-Raymont\inst{3} \and E. Ferrero\inst{2}}
\offprints{W. Brinkmann; e-mail: wpb@rzg.mpg.de}
\institute{Centre for Interdisciplinary Plasma Science,
 Max--Planck--Institut f\"ur extraterrestrische Physik,
 Postfach 1312, D-85741 Garching, FRG
\and Max--Planck--Institut f\"ur extraterrestrische Physik,
 Postfach 1312, D-85741 Garching, FRG
\and Mullard Space Science Laboratory, University College London,
             Holmbury St Mary, Dorking, Surrey, RH5 6NT, U.K.}
\date{Received ; accepted }

\abstract{In a $\sim$ 20~ksec XMM-Newton observation  the 
X-ray transient radio-quiet quasar \pg was found  in a 
historically high state compared to previous X-ray observations.
The quasar showed a featureless
spectrum with a strong soft excess over the extrapolation of 
a hard power law. Comptonization models or a broken power law with 
$\Gamma_{\mathrm {soft}} \sim 2.75$, $\Gamma_{\mathrm {hard}} \sim 2.25$
and a break energy of E$_{\mathrm {break}} \sim 1.35$ keV represent
acceptable descriptions of the spectral continuum. In the Comptonization 
models the temperature of the Comptonizing gas is considerably lower
than generally found in (broad line) Seyfert galaxies whereas the
 optical depth is
much higher. As a similar behavior has been seen in NLSy1 galaxies,
it might be an indicator of  the different  physical 
conditions in these two classes of AGN.
During the XMM-Newton observation the flux of  \pg 
varied achromatically  in a smooth, nearly linear fashion,
by $\sim 25$\% on time scales of a few 
thousand seconds, which puts some constraints on current models
of Comptonizing accretion disk coronae.
}

\titlerunning{XMM-Newton observations of PG 0844 + 349}
\authorrunning{W. Brinkmann et al. }
\maketitle

\section{Introduction}

Quasars as a class of Active Galactic Nuclei (AGN) exhibit a vast 
diversity  in their observed properties. 
Some of these, like the radio-loud - radio-quiet dichotomy, can be related
to the physical characteristics of the accreting central black hole and to 
morphological differences of the host galaxies.
Others can result from specific geometrical conditions of the accretion 
flow and from the particular  viewing condition of the observer with respect
to the quasar (see, for example, Elvis 2000).
In many cases, however, the causes for the unusual properties of a quasar remain
obscure and thus the study of objects displaying atypical characteristics 
might yield important clues to the physical processes governing 
the quasar emission.
 
One of the objects  with several extraordinary properties is PG~0844$+$349,
 a well
studied nearby (z=0.064), bright (m$_\mathrm{v}$=14.0), radio-quiet quasar from
the Palomar Green sample (Schmidt \& Green 1983). 
Deep radio observations reveal no nuclear source and
the faint detection in the field (S$_\mathrm{1.49~GHz} = 0.3$ mJy) might be related to an 
R = 20 mag object south-west of the quasar (Condon \etal 1987).
\pg is optically quite spectacular: it appears as a barred spiral
with complex outer structure and a nearly equally bright companion
galaxy (Hutchings \& Crampton 1990).
It belongs to the class of AGN with unusually strong Fe~II emission 
(Wang \etal 1996) and according to the optical spectrum displayed 
in Boroson \& Green (1992)
the source shows all features of a Narrow-Line Seyfert 1 galaxy (NLSy1)
 - strong FeII emission and weak forbidden narrow lines, although its
FWHM(H$\beta$) = 2420 km $\rm s^{-1}$ is slightly above the often used
cut-off line of 2000 km $\rm s^{-1}$ (e.g. Osterbrock \& Pogge, 1985).
However, this cut-off for the classification of NLSy1s is more artificial
than physical, so we still might define this source as a NLSy1 galaxy
especially as it shows X-ray properties typical for this class
of objects, i.e., a steep soft X-ray spectrum and strong variability
(for reviews see Boller \etal 2000). 
It is infrared-loud with a disturbed host galaxy (Clements 2000),
and it is strongly variable at all wavelengths on various time scales.
The optical flux seems to have changed by $\sim $70\%  between 1986 
(Elvis et al.
1994) and 1993 (Maoz et al. 1994) and intra-night variations with
amplitudes of $\Delta B = 0.08$ mag were detected by Jang et al. (1997). 
From reverberation measurements Kaspi et al. (2000) deduced a mass of
$\sim 2.5\times10^7$ M$_\odot$ for the quasar.
 
Compared to the ROSAT All Sky Survey (Yuan \etal 1998) the soft X-ray
flux of \pg had decreased by a factor of $\sim$6 in a pointed ROSAT
observation  6~months later, without any noticeable changes in the spectral
slope (Rachen et al. 1996).
During the X-ray low state \pg had an X-ray loudness of $\alpha_{\rm {ox}} > 2$, i.e.
it could be regarded as `X-ray weak' compared to the average 
$<\alpha_{\rm ox}> \sim 1.6$
for radio-quiet quasars (Yuan \etal 1998). 
In an ASCA observation the object was found  in a high state, with
a photon index of $\Gamma = 1.98$ and a Fe~K$\alpha$ line with EW $\sim$ 300 eV.
On shorter time scales, the X-ray flux in the 2--10 keV band is highly 
variable; the fastest
variation detected is 60\% in less than $2\times10^4$ s (Wang et al. 2000).
Historic light curves in the X-ray, UV, and optical bands indicate that the
variability amplitude in the UV and optical (up to 70\%)
 is much smaller than in X-rays (up to a factor of 10).
The optical micro-variability of this object can actually be driven by
re-processing of the variable X-ray flux if only half of the absorbed X-rays are
re-radiated in the optical-to-UV band.
The comparison with  Einstein (Kriss 1988)  and EXOSAT (Malaguti et al. 1994)
 observations  further 
showed that the quasar can be classified as X-ray weak only in one
out of five X-ray observations (for details see Wang \etal 2000). 

To what extent the intensity variations are related to spectral changes
is difficult to examine in detail due to the poor statistics of the 
pointed ROSAT/ASCA spectra, especially if significant X-ray scattering is 
present. Corbin \& Boroson (1996) mention that the quasar shows 
`associated absorption' in an HST spectrum  and  Wang et al. (2000) 
claim the existence of  narrow associated
Ly${\alpha}$ absorption. There are no HST data for the C IV region, but
co-added IUE data suggest C IV absorption consistent in velocity with
the Ly${\alpha}$ absorption.

\section {The XMM-Newton observation}

PG~0844$+$349 was observed by XMM-Newton on  November 5, 2000.
The EPIC PN was operated in Full Window mode with medium filter
with a total exposure of $\sim$ 21 ksec.
Both MOS cameras were in Large Window mode with medium 
filters as well, 
each yielding $\sim$ 23.5 ksec of data.
The two RGS chains were operated in Spectroscopy mode, both with exposures of
$\sim 26 $ ksec.
The OM camera was operated in Imaging mode and five images were taken
with the U-band filter (300--400 nm).   
For the data analysis we used the XMMSAS version 5.2.0 for EPIC and OM, and 
version 5.3.0 for RGS.

\subsection{The light curve}

We determined the 0.5$-$10~keV PN light curve of \pg by accumulating
 the photons in 400 sec bins
from a circular region of radius r = 100\arcsec ~around the quasar position; 
for this we selected single and double events with quality flag = 0
(for details of the EPIC detectors see Ehle \etal 2001).
For the MOS data we took singles and doubles (Pattern 0$-$12) as well
and selected the photons from a circular region of radius r = 60\arcsec ~ 
around the source.
 
\begin{figure}
\psfig{figure=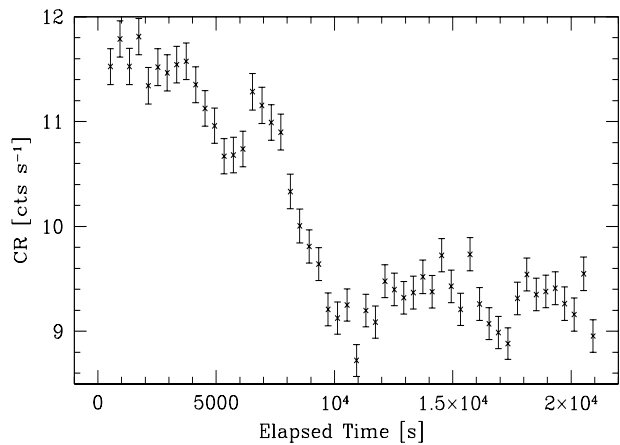,height=6.3truecm,width=8.3truecm,angle=0,%
 bbllx=51pt,bblly=259pt,bburx=555pt,bbury=599pt,clip=}
\caption[]{Background subtracted summed PN + MOS light curve of 
 PG 0844+349; the time binning is 400 sec.}
\label{figure:light}
\end{figure}

During the observation the background was rather noisy with some
 strong flares.
The total background count rate generally contributes only about
 4\% of the source count rate.
However, in two flares  the background count rate  at 
energies above 3 keV reaches up to 40\% of the source count rate
and care has to be taken in the spectral analysis.
In particular, we ignore data from the last 2~ksec of the observation as the
background count rate becomes very variable then.
Since there are no differences between PN and MOS light curves,
in order to increase the signal-to-noise ratio,  we summed up
the PN count rates (average 6.61 cts/s)
and the two MOS count rates (average MOS~1: 1.93 cts/s; MOS~2: 1.94 cts/s).
The summed, background subtracted, EPIC light curve shown in 
Fig.~\ref{figure:light}
starts at the switch on of the PN camera. There are about 2.5~ksec of MOS data
previous to that, where the source stays on the same high, flat count rate 
level.
The RGS observation started around the same time as the PN, and lasted some 
1.8 ksec longer, well into the period of flaring background.

The EPIC count rate (see Fig.~\ref{figure:light}) starts to decrease after 
about 4~ksec, recovers for
 $\sim$ 2~ksec and then changes by about 25\% 
 to a lower intensity level near the middle of the observation.
The flux drops occur in a smooth, nearly linear fashion 
at a rate of $\sim 6.1\times10^{-4}$ cts~s$^{-2}$.
We will show in the next section that these changes are achromatic to
high statistical significance.   
With the  spectral  information obtained below, the K-corrected 
0.5--10~keV luminosity during the first part of the observation is
$\sim 2.5\times10^{44}$ erg~s$^{-1}$.
We can use the $\sim$ 20\% decline of the count rate over $\sim$ 3500 s 
near the middle of the observation to estimate 
the lower limit of the radiative efficiency,
$\eta > 5\times10^{-43} \Delta L / \Delta t$ (Fabian 1984), and we
obtain $\eta > 7\times10^{-3}$ which is far below the theoretical limit
for a Schwarzschild black hole.
 
The OM provided U-filter images integrated for $\sim $ 4000~s each.
The brightness of the source, obtained from the Pipeline Processing
System (PPS), stayed approximately constant at 14.00 mag; the largest
deviation from this average value occured during the period 
$\sim$ 2--6~ksec after the start of the
PN exposure, during the first intensity dip,
 when the source was 0.01 mag fainter than the average. This
is, however, only a $\sim 1.5 \sigma$ effect which implies that
any variations of the optical flux are negligible compared to the 
changes in the X-ray band. 
  
\begin{figure}
\psfig{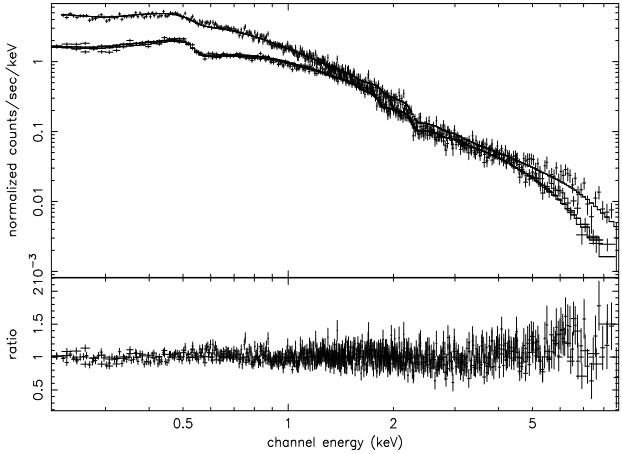}
\caption[]{Broken power law fit to the total PN + combined MOS data of 
    PG 0844+349; the PN spectrum is the upper curve. }
\label{figure:all4fit}
\end{figure}

\subsection{Spectral properties}
	
For the EPIC spectral analysis we used the latest available response matrices
(version 6.1) issued in December 2001.
For the PN camera we extracted single and double events with quality flag = 0
from a circular region with a radius of 45\arcsec ~around the source
position. 
This radius includes $\sim$ 88\% of the source photons (Ghizzardi \& Molendi 2001)
but avoids the gap between the detector chips.
The background was taken on the same chip at a similar offset position.
With a count rate of $\ma$ 7 cts~s$^{-1}$ in the high state the PN detector, 
operated in Full Window mode, showed strong indications of pile-up, clearly 
apparent from the XMMSAS task $epatplot$. 
We therefore discarded photons from the innermost $3\times3$ RAW pixels 
at the core of the point spread function from the spectral analysis.
As a consequence, the normalizations of the models in the
PN fits given in Table 1 are underestimating the actual photon flux. 
Further, we discarded the data of the last $\sim$ 2 ksec of the observation,
when the background was strongly flaring, and we excluded times of high background, 
i.e. greater than 50 cts/s, from the Good Time Intervals (GTI) file.
However, this last 
selection criterion influenced the quality of the fits only marginally. 
For the two MOS cameras we selected events with  pattern $\leq 12$ and flag = 0
and a similar extraction radius as for the PN;
 the background was taken from an area on the 
same chips near the source.  Pile-up did not influence the spectral fits
of the MOS data noticeably; therefore the full data were used.
 
The RGS spectra of \pg were accumulated by selecting counts from a 
region centered on the source position and enclosing 90\% of the point 
spread function in the cross dispersion direction, and making the first order 
selection by including 90\% of the expected CCD pulse height 
distribution of the source photons. The background spectrum was obtained 
from two spatial regions excluding 95\% of the source point spread function
in the cross dispersion direction and performing the same order selection
as for the source.

While the ASCA high state data could be well fitted by a simple absorbed 
power law (Wang et al. 2000), 
this model does not provide an acceptable fit to the PN 
 data from the first, high state part of the observation. 
With a reduced $\chi^2 = 1.42$ the fitted power law index 
is $\Gamma = 2.57$  and the residuals clearly show that at higher energies the
 spectrum gets flatter.
A broken power law, fixing the absorption at the Galactic value 
N$_{\mathrm H} = 3.32\times10^{20}$cm$^{-2}$ (Lockman \& Savage 1995),  
gives an acceptable fit to the PN data
 with a reduced $\chi^2 = 1.03$ for 372 d.o.f. (see Table 1).
Fitting the same time interval with the MOS1 and the MOS 2 data separately 
yielded  similar acceptable fits.
The fit parameters in Table 1 indicate that the two MOS spectra have nearly 
identical power law slopes.
In both MOS cameras the residuals
at the lowest energies show an S-shaped pattern with a maximum at $\sim 0.5$ keV, 
while the PN fits, with generally slightly flatter power laws and a higher
break energy, show 
systematic positive residuals at $\sim 0.6$ keV.

Applying the same model to the low state, i.e. the period between 10 and
19 ksec after the start of the observation, gave an acceptable fit with
nearly identical parameters, with the same `typical' differences
between the three detectors as above (see Table 1). 

\begin{table*}
\small
\tabcolsep1ex
\caption{\label{fits} Results from spectral fitting assuming fixed Galactic 
N$_{\mathrm H} = 3.32\times10^{20}$cm$^{-2}$.}
\begin{tabular}{lclcclcl}
\noalign{\smallskip} \hline \noalign{\smallskip}
\multicolumn{1}{c}{Period} & \multicolumn{1}{c}{Detector} &
\multicolumn{1}{c}{Model} & \multicolumn{1}{c}{$\Gamma_{\rm soft}$} &
\multicolumn{1}{c}{E$_{\rm break}$} & \multicolumn{1}{c}{$\Gamma_{\rm hard}$}
& \multicolumn{1}{c}{Power law norm.} 
 & \multicolumn{1}{c}{$\chi^2_{\rm red}$/dof} \\
\multicolumn{1}{c}{ } & \multicolumn{1}{c}{  } & \multicolumn{1}{c}{   }
&\multicolumn{1}{c}{ } & \multicolumn{1}{c}{(keV)} & \multicolumn{1}{c}{ } &
\multicolumn{1}{c}{(10$^{-3}$ ph/keV/cm$^2$/s) } & \multicolumn{1}{c}{}\\
\noalign{\smallskip} \hline \noalign{\smallskip}
High & PN & bknpow &2.71$\pm0.02$ & 1.63$\pm0.12$ &  2.13$\pm0.06$ & 
   1.65$\pm$0.03 &1.03/372\\
  &  MOS 1 & bknpow &2.81$\pm0.04$ & 1.02$\pm0.08$ &  2.31$\pm0.04$ &
   2.96$\pm$0.12 &  1.03/148\\
  & MOS 2 & bknpow &2.85$\pm0.04$ & 1.03$\pm0.08$ &  2.31$\pm0.04$ &
   3.06$\pm$0.12 & 1.10/145\\
Low & PN & bknpow &2.75$\pm0.02$ & 1.81$\pm0.11$ &  2.08$\pm0.06$ &
   1.32$\pm$0.02 & 1.04/372\\
  & MOS 1 & bknpow &2.77$\pm0.03$ & 1.26$\pm0.08$ &  2.27$\pm0.04$ &
   2.49$\pm$0.06 &  0.92/167\\
  & MOS 2 & bknpow &2.93$\pm0.03$ & 1.05$\pm0.06$ &  2.34$\pm0.03$ &
   2.45$\pm$0.08 & 1.02/166\\
Total & PN & bknpow & 2.72$\pm0.01$ & 1.80$\pm0.08$ & 2.09$\pm0.04$ &
   1.46$\pm$0.02 &  0.98/537\\
 & PN + MOS 1+2 & bknpow & 2.77$\pm0.09$ & 1.33$\pm0.03$ & 2.25$\pm0.02$ &
   (a) & 1.11/1006\\
Total$^{(1)}$ & PN & pow &  &  & 2.05$\pm 0.05$  & 0.96$\pm$0.10 & 0.83/188\\
 &  PN & pow + gauss & 2.11$\pm 0.05$ & 6.247$\pm0.127^{(2)}$ & 0.24$\pm0.16^{(3)}$&
   0.96$\pm$0.10 &0.81/185\\ 
Total & PN & pow + bbody &  & 1.94$\pm0.14^{(4)}$ & 2.73$\pm0.01$&
   1.45$\pm$0.02 & 0.99/629\\
      & PN & pow + brems &  & 0.29$\pm0.01^{(4)}$ & 2.11$\pm0.03$&
   1.02$\pm$0.06 & 0.95/621\\
      & PN & pow + diskbb & & 0.14$\pm0.01^{(4)}$ & 2.23$\pm0.02$ &
   1.17$\pm$0.04 & 1.00/629\\
      & PN & pow + mekal & & 0.26$\pm0.01^{(4)}$ & 2.18$\pm0.03$ &
   1.12$\pm$0.05 & 0.95/627\\
\noalign{\smallskip}\hline

\end{tabular}
\medskip

(a): Normalizations for the individual detectors allowed to be different.\\
(1): Fit over the 2$-$9 keV energy band only; (2) Fitted line energy; (3):
 Line width (sigma) in keV; the photon flux in the line is \\ 
 (5.7$\pm4.9)\times10^{-6}$ photons/cm$^2$/s. 
(4): Temperature of the additional model component in keV. \\
%(5): RGS energy band 0.3--2.1 keV (6--38 \AA). \\

\end{table*}

We repeated the fit over several other, shorter time intervals during 
the observation and always
obtained acceptable fits with very similar parameters.
We thus conclude that no statistically significant spectral variations
occur during the flux changes of the source. 
Therefore for the subsequent fits we used the data from the full observation,
again excluding the last 2~ksec and times of high background.
 
Fig.\ref{figure:all4fit} shows the simultaneous broken power law fit to the 
whole data set for the PN and the combined MOS
instruments. Due to the slight differences between the PN and MOS cameras,
visible at $\sim$ 0.5--0.6 keV and at higher energies, the quality of the
fit is only moderate ($\chi^2_{\mathrm {red}} = 1.11$ for 1006 d.o.f.).
Adding a Gaussian iron line does not improve the fit, mainly because of the low
statistical significance of the data at higher energies.
We therefore fitted the PN data separately for the 2--9 keV range with a
single power law and, secondly, with a power law plus a Gaussian line.
We find a broad line at an  energy of 6.25$\pm$0.13 keV 
(E$_{\rm l,rest} = 6.65 \pm 0.13$ 
keV in the quasar's rest frame) and an equivalent width of $\sim$ 235 eV. 
For a narrow line  (i.e. fixing the Gaussian sigma at 10~eV) we get the 
same line energy and an equivalent width of $\sim$ 120 eV. 
Including the lines improves the quality of the fits
marginally (see Table 1),
but only at a $\sim$ 94\% confidence level, according to an F-test. 
The small equivalent width found is atypical for NLSy1 galaxies, for
which BeppoSax (Comastri 2000) and ASCA (Turner \etal  1998) generally
find ionized iron lines with large equivalent widths (few 100 eV). 
 
The moderate quality of the combined PN and MOS fits and the fit residuals 
can be explained
as resulting from the remaining uncertainties in the detector calibrations.
However, they might also indicate that the chosen spectral model  
can be improved.
  
\begin{figure}
\psfig{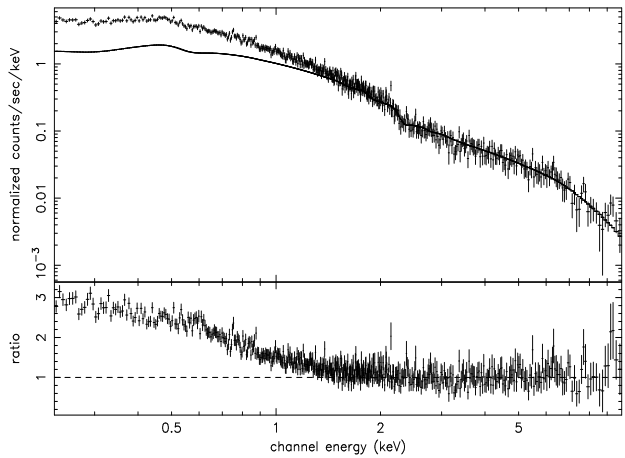}
\caption[]{Power law fit to the total PN data in the energy range 2$-$10 keV;
    the fitted model is extrapolated to lower energies.
    The lower panel gives the ratio between data and model. }
\label{figure:pnfit}
\end{figure}

In Fig. \ref{figure:pnfit} we show the power law fit to the total PN data 
for the 2$-$10 keV range, extrapolated to lower energies.
The slope of $\Gamma \sim 2.04$, consistent with ASCA, is probably not
affected by contributions of more complex models for the low energy
part of the spectrum.
 The ratio between the data and the model
clearly shows the  `gradual soft excess' (Pounds \& Reeves 2002) 
also found in other AGN with the high sensitivity and large bandwidth of 
XMM-Newton.
While the hard power law tail seems to be well constrained, the exact
shape of the spectrum at low energies remains uncertain.
We therefore tried several composite models, mainly to characterize the 
softer part of the spectrum. 
  
Models including a thermal component in addition to the hard power
law fit the data quite well, as shown in Table 1; the relative contribution of
the power law component can be inferred from the listed normalization.
 The power law + bbody fit provides two solutions: one in which the power 
 law component represents the hard flux, the black body temperature is 
 similar to that of the diskbbody and another, perhaps less physical
 solution, where the steep power law provides the soft flux, the hot
 bbody the hard flux (these parameters are given in Table 1).  
All models are statistically acceptable (although the  above mentioned 
systematic residuals at $\sim$ 0.6 keV persist in all 
fits) and therefore the physical nature of the soft emission remains unclear.

As the fitted slope of the spectrum at higher energies seems to harden
with energy we tried to fit a curved continuum model (Fossati \etal 2000)
but no good fit could be achieved.
We further tried several composite cases fixing
the hard power law index at the best fit slope of the 2$-$10 keV fit
but we never obtained an acceptable fit: this seems to indicate that 
the hard power law is only the tail of a distribution and not a 
separate component.

\begin{figure}
\psfig{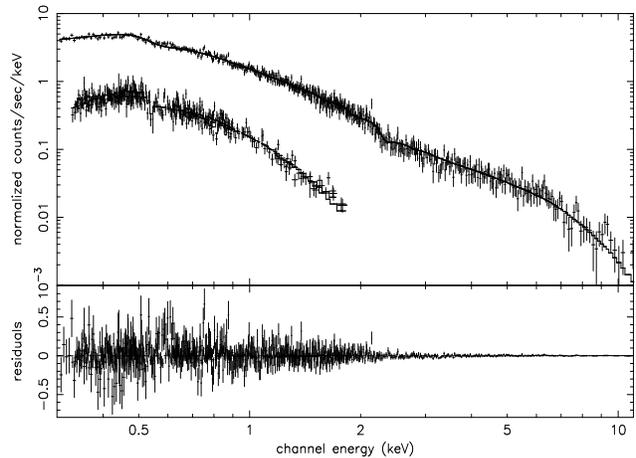}
\caption[]{Power law + bremsstrahlung fit to the PN, RGS1 and RGS2 
data of PG 0844+349.}
\label{figure:rgsfit}
\end{figure}

Nearly featureless spectra with a strong soft component are indicative
of Comptonization of soft photons in the hot corona of an accretion disk
(Haardt \& Maraschi 1993).
A fit with the Xspec Comptonization model $comptt$ (Titarchuk 1994) 
failed to
reproduce the hard power law and resulted in an unacceptable fit 
($\chi^2_{\mathrm {red}} = 1.97$).
Adding a hard power law to the $comptt$ model
($\chi^2_{\rm red} = 1.09 / 1094$ d.o.f) or fitting the sum of two 
Comptonization models with different temperatures and optical depths of the
scattering medium ($\chi^2_{\mathrm {red}} = 1.11 / 1093$ d.o.f) 
resulted in fits nearly as good as the models in Table 1.
For the double - $comptt$ fit we either assumed the same temperature 
for the two soft photon components or left them free to vary independently.
In all three cases the bulk of the flux up to $\sim$ 3 keV originates from 
Comptonization of soft photons off a gas of kT $\sim$ 4.5 keV and
optical depth $\tau \sim 2.4$. The high energy part of the spectrum 
is formed either by the power law component or the second Comptonization
component with kT$\sim$ 16 keV and $\tau \sim 3.3$. 
The temperatures of the soft photons were fitted to be 
 kT$_0 = 65^{+99}_{-65}$
eV in case of the extra power law component, kT$_0 = 3^{+71}_{-3}$ eV 
(using the same values for the two soft components) and  
kT$_0(1) = 12^{+169}_{-12}$ eV and 
kT$_0(2) = 2^{+341}_{-2}$ eV  for the case of two independent 
soft components, respectively.
The temperatures of the Comptonizing electrons 
are at the low end of the expected temperature range
for AGN (Haardt \& Maraschi 1993) while the fitted optical depths are 
higher than usually deduced.  The temperatures of the soft photons 
appear rather low; however,
the parameters are very poorly constrained because of the close coupling
of temperature and optical depth in the Comptonization models.
Interestingly, the high optical depth implies that any reflection 
features (like the iron line) are suppressed by Compton scattering in the
corona itself (Matt \etal 1997).

The superior energy resolution of the RGS data could provide an ultimate
test for the nature of the soft emission. The same composite models 
tried on the PN data alone (see Table 1) were fitted simultaneously 
to the RGS1 and RGS2 spectra, as well as to the PN. The hard power law
slope and the temperature of the soft component were kept fix at the best 
fit value obtained from the fits of the PN alone, and the normalizations
for the three instruments were allowed to vary independently of each other
(because of the PN data selection used to avoid pile-up). The low energy 
absorption column was fixed at the Galactic value. 
Values of $\chi^2_{\mathrm {red}}$/dof very similar to those in Table 1 for the PN alone
were obtained (in the range 0.98--1.05/841),
with the formally best fit model being the combination of a power law and a
bremsstrahlung component. This best fit 
and the data are shown in Fig.\ref{figure:rgsfit}; the fit of the 
pow/diskbb and pow/mekal are virtually indistinguishable from this.
 Although the signal-to-noise ratio of the RGS data is only moderate, it is 
clear that there is no evidence of emission or absorption 
structures in the spectrum, indicating  the absence of any additional
source intrinsic absorbers. In particular, there is no evidence of excess
emission in the RGS at $\sim$0.6 keV which could explain the amplitude and the
shape of the residuals observed in the PN fit.

In the following, we use the broken power law model for the determination
of the source's energetics; however, it should be kept in mind that this is
a rather artificial representation of the source's spectrum and that at the 
lowest energies
the flux distribution might be quite different from a simple power law,
 introducing some uncertainty in the numerical values.

The observed average un-absorbed flux in the high state in the 
0.2--10.0 keV range amounts to 
2.6$\times10^{-11}$erg~cm$^{-2}$s$^{-1}$, while the average in the low state is
2.1$\times10^{-11}$erg~cm$^{-2}$s$^{-1}$.
With these fluxes the K-corrected X-ray luminosity of \pg is in the range
 L$_{0.2-10~{\mathrm {keV}}} = (4.1 - 4.9)\times10^{44}$ erg~s$^{-1}$, using a Friedman
cosmology with H$_0$ = 50 km~s$^{-1}$ Mpc$^{-1}$ and q$_0$ = 0.5.
The average soft-band luminosity is L$_{0.5-2~{\mathrm {keV}}} \sim 6.4\times10^{43}$
 erg~s$^{-1}$, for the hard band we find 
 L$_{2-10~{\mathrm {keV}}} \sim 6\times10^{43}$ erg~s$^{-1}$. 
Thus \pg is a quasar of rather low luminosity in the X-ray band.
Using the monochromatic luminosity at 2~keV we derive an X-ray loudness
$\alpha_{\rm ox} =  -0.384 \log (L_{2~{\mathrm {keV}}} / L_{2500\AA}) \sim 1.58$,
which is typical for average radio-quiet quasars (Yuan \etal 1998).

\section{Discussion}
 
Using the normalizations of the  power law fits
the above analysis shows that \pg was in a historically
high state (see Table 1 of  Wang \etal 2000) during
the XMM-Newton observation, with an average  0.1$-$2.4~keV luminosity
of $\la 3 \times10^{44}$ erg~s$^{-1}$.
Judging from the OM data the object was optically rather bright as well.
An extrapolation of the soft power law into the optical band over-predicts
the optical flux by a factor of $\sim 4$, therefore the spectrum must
break between the optical and the soft X-ray band. 
Nevertheless, most of the power is emitted in the soft X-ray / UV band.
Taking the mass estimates for the central black hole from Kaspi et al. (2000)
(2$-$3$\times10^7$M$_\odot$) 
this implies that \pg must be a quite efficient accretor.
The optical spectrum of \pg resembles very much that of NLSy1 galaxies
(Boroson \& Green 1992 ) and thus an efficient accretion flow  as well as its
X-ray variability would be in accordance
with current models of  these AGN.

The X-ray spectrum of \pg shows a relatively flat power law component at
higher energies ($\Gamma_{\rm hard} \sim 2.25$) and a strong, steep  
soft excess at energies $\la$2 keV. 
Both components are featureless and the inclusion of a broad  or a
narrow iron line 
around 6.7 keV improves the fit only marginally.
The soft excess can be modeled with various components resulting in
fits with similar statistical significances. The RGS data do not
indicate the presence of any spectral feature either. 
The excess of the soft flux reaches a factor of $\sim$2.5 over the
extrapolation of the hard power law.
 
This kind of spectrum is not unusual amongst recent measurements
of X-ray spectra of bright Seyferts / quasars with XMM-Newton 
(O'Brien \etal 2001, Page \etal 2001, Pounds \& Reeves 2002).  
As a matter of fact, the spectrum of \pg appears to be  a carbon copy
of that of  the  NLSy1 galaxy PKS~0558-504 (Pounds \& Reeves 2002,
 Fig. 1), even with 
respect to the `big blue bump' seen in both objects (O`Brien \etal 2001).
 However, we also note that the power law slope of \pg is slightly steeper 
 than that of PKS~0558-504, which is radio-loud and shows correlations
between X-ray brightness and hardness (Gliozzi et al. 2001), 
characteristics not shared by \pg.

The featureless spectra in these objects indicate that we are seeing the 
bare continuum disk emission from the quasars: thus disk Comptonization models,
where the X-rays are produced via inverse Compton emission in a hot
corona embedding a cooler accretion disk 
(e.g. Haardt \& Maraschi 1993, Pounds \etal 1995),
might provide a satisfactory physical explanation. 
Fits with available models for this scenario, however, yield 
parameters for \pg not typically found in other AGN, and not statistically  
preferred according to the above analysis: this might indicate that the
actual physical conditions in the sources are more complex than our
simplified models can account for.  
 
The large variety of possible Comptonization scenarios (for example, 
Haardt 1996, Zycki \etal 2001) does not allow a better confinement of the 
parameters in the physical phase space.
Stronger constraints might be deduceable from an analysis of the temporal
behavior of the source. PG 0844+349 was known to be variable in X-rays from
previous ROSAT and ASCA observations, but only XMM-Newton is able to follow
the flux changes on the shortest time scales.
The X-ray flux  (Fig. \ref{figure:light}) basically changes 
during the observation 
from a  higher to a lower level in a nearly linear fashion  and  the 
slopes of the intensity variations are very similar, indicating a
rather well organized process. 
The light curve shows (for a NLSy1) an atypically low variance  and the 
intensity changes occur without any measurable spectral changes in the X-ray band. 
  
The optical image of \pg indicates that we are seeing the object nearly pole-on
and therefore heavy obscuration of the X-ray emitting region appears unlikely.
The emission from a jet, changing its geometrical appearance, can very likely
be ruled out as the source is definitely radio-quiet, unless we are
looking directly into a purely  hydrodynamically outflowing jet for which 
there are no other observational indications. 

Achromatic flux changes can result from  changes of the effective radiating 
area, ensuring that the spectral shape does not change noticeably. While the 
temperature of the cold matter does not play a significant role in 
the models, changes in the optical depth $\tau$ give rise to significant 
spectral variability (Haardt \etal 1997). Most of the theoretical
investigations are, however, concentrating on the higher energy
part of the AGN spectrum (see, for example, Petrucci \etal 2001), and the
parameter space  of relevance here is only poorly explored.
In any case, changes of the radiating area  
are expected to happen on the dynamical time scale for Keplerian 
inflow, $\tau \sim 9~ 10^3~  (\mathrm{r/R}_{\mathrm s})^{3/2}~ 
({\mathrm M}_{\rm bh}/10^7 ~ {\mathrm M}_\odot) $ sec, where 
R$_{\mathrm s}$ is the Schwarzschild radius of the central black hole
 of mass M$_{\rm bh}$. 
This estimate implies a rather low mass for the central black hole in
\pg and a compact emitting region. However, strong variability of 
Seyferts and Galactic black hole candidates indicates that the corona 
cannot be a uniform, continuous medium, unless it is geometrically
thin (Celotti \etal 1992); moreover,  observational evidence implies 
that the geometry of the coronal plasma cannot be slab like, but is made
up of a number of distinct active regions (Haardt \etal 1994).
 
In a popular class of models the corona is heated by magnetic
fields which rise up buoyantly from the disk where they reconnect and release
 their energy in flares. The energy storage in the corona is very likely
the magnetic fields (Merloni \& Fabian 2001) and the picture of the 
corona is that of a spread of active regions, of which only a few are 
large and dominate at any given time.
The overall time scale of the evolution of the magnetic field 
configuration is expected to be of the order of the Keplerian time 
scale again (Romanova \etal 1998) but individual active regions 
can certainly evolve much more rapidly, depending of the magnetic
field configuration and the flow conditions in the disk, which are
largely unknown.  

\section{Conclusions}

In a 20 ksec XMM-Newton observation  the X-ray transient quasar  \pg 
was found in a historically high X-ray state with an average 0.2$-$10 keV
luminosity of L$_{0.2-10~{\mathrm {keV}}} \sim 4.5 \times10^{44}$ erg~s$^{-1}$.
During the observation the flux of  \pg 
varied achromatically by $\sim 25$\% from a higher into a lower
flux state in a very smooth manner on a time scale of a few 
thousand seconds.
  
The quasar showed a featureless
spectrum which can be physically explained by Comptonization from hot
electrons of the emission of an accretion disk.
The rather low signal-to-noise ratio RGS spectra do not indicate the 
presence of any emission or absorption features either.
The fitted values of the slopes of the power law type spectrum at high 
energies ($\Gamma_{\mathrm {hard}} \sim 2.05$) and those of the power law in
the soft band 
($\Gamma_{\mathrm {soft}} \sim 2.75$) differ slightly between the 
EPIC detectors and depend on the complexity of the models fitted to the data.
Current Comptonization models predict considerably lower temperatures and 
higher optical depths of the Compton scattering electrons than generally
 found in Seyfert galaxies and radio-quiet quasars (see e.g. 
Petrucci \etal 2001).
As \pg shares many of the characteristics of NLSy1 galaxies and similar
parameters were recently deduced for other NLSy1 galaxies 
(O'Brien \etal 2001, Page \etal 2001) we  propose that Comptonization
spectra with high optical depths and moderate plasma temperatures
might be a distinguishing criterion for the accretion 
process in NLSy1 galaxies. 
 
More sensitive XMM-Newton observations and a deeper theoretical exploration
of this lower energy and lower temperature parameter space of
Comptonization models might shed some new light into the physical
properties of these objects.

\vskip 0.3cm
\begin{acknowledgements}
This research has made use of the NASA/IPAC Extragalactic Data Base
(NED) which is operated by the Jet Propulsion Laboratory, California
Institute of Technology, under contract with the National Aeronautics
and Space Administration.
This work is based on observations with XMM-Newton, an ESA science mission
with instruments and contributions directly funded by ESA Member States and the
USA (NASA). The Mullard Space Science Laboratory acknowledges financial 
support from the UK Particle Physics and Astronomy Research Council.
\end{acknowledgements}

%\special{!userdict begin /bop-hook{gsave 300 600 translate
%20 rotate /Times-Roman findfont 80 scalefont setfont
%0 0 moveto 0.9 setgray (02.05.02) show grestore}def end}
\end{document}